\documentclass[%
superscriptaddress,
groupedaddress,
nofootinbib,
 amsmath,amssymb,
 aps,
prc,
]{revtex4-2}

\usepackage{graphicx}
\usepackage{dcolumn}
\usepackage{bm}

\usepackage{color}
\usepackage{caption,subcaption,float}

\usepackage[dvipsnames]{xcolor}
\newcommand{\myColor}{OliveGreen}

\usepackage[colorlinks=true,urlcolor=\myColor,citecolor=\myColor,linkcolor=\myColor]{hyperref}
\usepackage{cleveref}
\usepackage{nicematrix}

\newcommand{\pbpb}{$Pb-Pb$ }
\newcommand{\ppb}{$p-Pb$ }
\newcommand{\auau}{$Au-Au$ }
\newcommand{\dau}{$d-Au$ }
\newcommand{\heau}{$He-Au$ }
\renewcommand{\aa}{$AA$ }
\newcommand{\pp}{$pp$ }

\newcommand{\highpt}{high-$p_T$ }
\newcommand{\lowpt}{low-$p_T$ }
\newcommand{\raa}{$R_{AA}$ }
\newcommand{\rpa}{$R_{pA}$ }

\newcommand{\jewel}{\textsc{Jewel}}

\begin{document}

\preprint{APS/123-QED}

    \title{JEWEL on a (2+1)D background}
    \title{JEWEL on a (2+1)D background with applications to small systems and substructure}
    
    \author{Isobel Kolbé}
    \email{Email: isobel.kolbe@usc.es}
    \affiliation{Instituto Galego de Fisica de Altas Enerxias (IGFAE), Universidade de Santiago de Compostela, E-15782 Galicia, Spain}

    \date{\today}
    
    \begin{abstract}
        High-$p_T$ jets are an important tool for characterizing the quark-gluon plasma (QGP) created in heavy-ion collisions. 
        However, a precise understanding of the jet-medium interaction is still lacking, and the development of more sophisticated observables is needed. 
        This work presents a tool that allows for the exploration of alternative high-$p_T$ observables in a variety of collision systems. 
        The tool builds on the publicly available JEWEL Monte Carlo code, and allows for the evolution of a jet on any given (2+1)-dimensional background. 
        Proof-of-concept observables are also presented, studied using the latest version of JEWEL, JEWEL-2.4.0. 
        The simplicity of the separation of the physics of the jet from the physics of the medium (while still allowing for the usual JEWEL medium-response), allows for easy interpretation without the need for complex parameterization. 
        Results are produced using the RIVET toolkit, which allows for transparent preservation and development of analyses that are compatible with experimental methods. 
        The code and analysis used to produce the plots presented here are made publicly available on a Github repository, with up-to-date usage instructions. 
        This tool is expected to be useful to the broad jet-physics community for the study of precision observables for jets.
    \end{abstract}
    
    \maketitle
    

    
    \section{\label{sec:Introduction}Introduction}
    
        The modification of \highpt jets in heavy-ion collisions is a critical signature of the hot and dense matter created in heavy-ion collisions.
        In central heavy-ion collisions, the main channel for the modification of \highpt jets is known as ``jet suppression'', which has enjoyed both theoretical and experimental success (see \cite{Apolinario:2022vzg} for a recent review).
        However, while the theoretical understanding of the jet-medium interaction has become very sophisticated, a need has arisen for more precise observables.
     
        This paper presents a tool that may be used to expolore alternative \highpt observables, specifically in a variety of collision systems.
        This work builds on an existing jet Monte Carlo (MC) code, the publicly available \jewel\, \cite{Zapp:2012ak,Zapp:2013vla,Zapp:2008gi}, providing \jewel\, with the ability to evolve a jet on any given (2+1)-dimensional background.
        Presented here are also proof-of-concept observables studied using the latest version of \jewel \,,\jewel-2.4.0.
        Although more sophisticated jet event generators exist and are often publicly available, the value of the present interface lies in its simplicity.
        The conceptual separation of the physics of the jet, handled by \jewel\,, and the physics of the medium, allows for simple interpretation that remains unclouded by the subtle interplay between various model parameters and effects.
    
        The results presented here are produced using version 3 of the RIVET toolkit \cite{Bierlich:2019rhm}.
        Within RIVET, jets are clustered using the FASTJET package \cite{Cacciari:2005hq,Cacciari:2011ma}, as well as the LundPlane extension based on \cite{Dreyer:2018nbf}.
        Additionally, the \jewel-2.4.0 release is accompanied by a RIVET projection for the constituent subtraction method.
        The public availability of RIVET and its design philosophy of transparent preservation and development of analyses that are compatable with experimental methods means it is useful to develop observables with RIVET analyses.

        Since it is hoped that this tool will be useful to the broad jet-physics community, the code is made publicly available, along with the RIVET analysis used to produce the plots presented here.
        The code is hosted on a Github repository \href{https://github.com/isobelkolbe/jewel-2.4.0-2D.git}{https://github.com/isobelkolbe/jewel-2.4.0-2D.git}, along with upt-to-date usage instructions.
    
        This paper is organised as follows:
        In \cref{sec:Software}, the medium interface for \jewel\, is presented with a brief discussion of its features and validation.
        In \cref{sec:Substructure}, the problem of small systems is explored by computing illustrative substructure examples using the (2+1)D medium interface.

    \section{Software}\label{sec:Software}

        \subsection{Basic features}
        
            Of particular importance is the new subtraction method (constituent subtraction)\cite{Milhano:2022kzx}, which improves \jewel's reproduction of the jet mass and will be particularly important for other jet substructure observables.
    
            In the standard release, \jewel-2.4.0 can be run in two modes: in vacuum (hereinafter ``vac''), and with a simple medium model (hereinafter ``medium''). 
            The simple medium is a radially symmetric, longitudinally expanding temperature profile whose initial state is determined by a Glauber overlap of Woods-Saxon thickness functions and a given initial temperature.
            \jewel-2.4.0 has also been upgraded to use LHAPDF 6 \cite{Buckley:2014ana} and therefore has access to a wide range of nuclear Parton Distribution Functions (nPDFs).
    
            In order to study jets in a variety of collision systems, it is useful to be able to evolve the jet on a changing background.
            A natural choice for the determination of the properties of such a background is a hydrodynamic simulation, but it is also useful to be able to use any arbitrary background.
             \jewel\, has been used before in conjunction with a hydrodynamic background, particularly to study the effect of jets on the source terms in hydrodynamic simulations \cite{Floerchinger:2014yqa,Zapp:2014msa}.
            In addition to these studies, hydrodynamic plugins for \jewel\, have also been developed to study the sensitive interplay between elliptic flow and \raa \cite{Barreto:2022ulg,Canedo:2020xzf,Barreto:2021fbt}.
            However, to the author's knowledge, such plugins have not yet been used to study jet substructure in small systems, nor are they public. 
      
            The present work is built heavily on an interface originally implemented by Korinna Zapp, one of the authors of \cite{Floerchinger:2014yqa,Zapp:2014msa}, which is in turn similar in spirit to the \texttt{medium-simple.f} medium interface that is part of the standard \jewel\, release.
            The plugin presented here, named simply `\texttt{medium-2D}` so that, when built with \jewel\, will result in an executable \texttt{jewel-2.4.0-2D}, has the following features: The interface
                \begin{enumerate}
                    \item reads in temperature and velocity contours for up to 90 time steps.
                    \item can optionally read the locations of the binary collisions that source the background generation.  This is used to sample the location of the initial hard scattering (the default is to sample Woods-Saxon overlap functions).
                    \item correctly boosts into the rest frame of the fluid cell in order to determine the fluid density.
                    \item interfaces with the main \jewel\, code.
                    \item (as with the standard \jewel\, release), \textit{is not} able to use two different parton distribution functions (PDFs) in an assymmetric system.  Care must be taken to ensure that observables are either not sensitive to the PDF, or that the sensitivity to the PDF is carefully taken into account.
                    \item does not output the hadronized products of the medium to the \jewel\, event record: as in the standard \jewel\, release, the event record is a superposition of the \jewel-evolved jet and a PYTHIA-produced event.
                \end{enumerate}

            The medium interface can be passed a parameter file by passing the name of the parameter file to the \texttt{MEDIUMPARAMS} parameter in the \jewel.
            The interface can take many of the same parameters as \jewel's simple medium since it must still create an initial condition from which to sample the location of the initial hard scattering if the user does not provide an $N_{coll}$ probability density.

            Updated instructions for use are maintained on the code repository.

        \subsection{Testing and Validation}

                \begin{figure}
                    \centering
                    \includegraphics[scale=0.8]{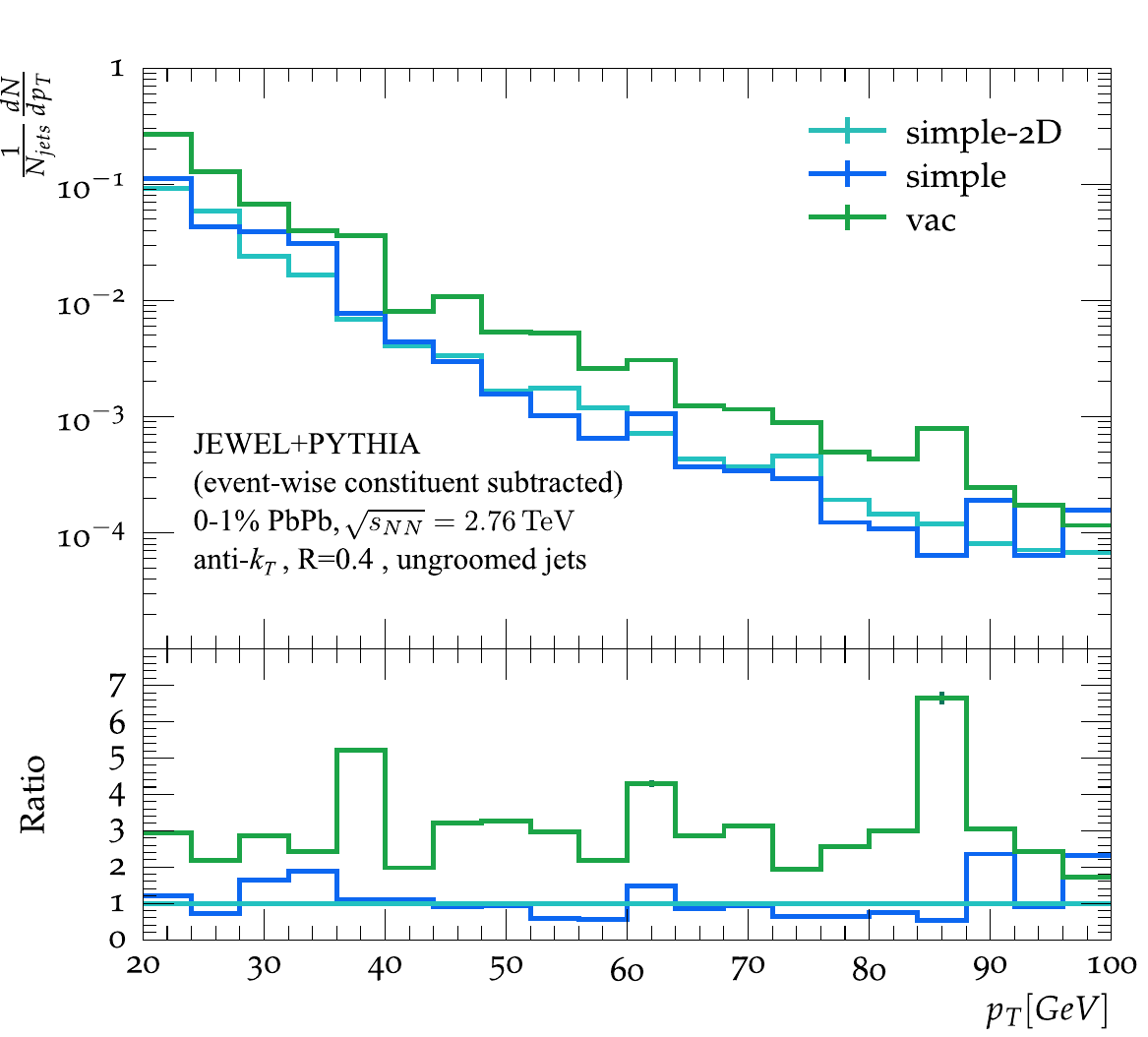}
                    \caption{The jet yield as a function of $p_T$ for jets evolved in \jewel, either without a medium (green, ``vac''), or with the same simple medium using either \jewel's own ``simple'' mode (cyan, ``simple''), or using the 2D interface presented here with medium profiles generated with \jewel's simple medium code (blue, ``simple-2D'') .
                    Monte Carlo errors are smaller than the thickness of the line. }
                    \label{fig:pt_none_edited}
                \end{figure}

            In order to test the interface, a set of temperature contours were produced using the code from \jewel's own \texttt{medium-simple.f} with no modifications.
            \jewel\, was then run using the 2D interface presented here with these contours (along with null velocity contours) as input, and the resulting distributions compared.
            As an illustrative example, the jet yield as a function of transverse momentum is shown in \cref{fig:pt_none_edited} for `simple-2D', `simple', and `vac', corresponding to running \jewel\, with the present 2D interface on medium profiles produced using \jewel's internal simple medium code, with the standard simple medium, and in vacuum, respectively.
            As expected, the vacuum case deviates significantly from the two medium cases.
            The systematic differences between ``simple'' and ``simple-2D'' are very sensitive to the granularity of the temperature grid.
            While this can be improved significantly, the memory cost is high.

    \section{Jet Substructure and Small Systems}\label{sec:Substructure}

        The medium interface presented here has broad applicability with in the study of jets in heavy-ion collisions.
        A particular concern within the community at present is related to small colliding systems such as proton-lead and (proton, deuteron, helium)-gold.
        In this section, the problem of small systems is explored.
        An argument is made for the need to develop substructure observables that are independent of the yield but sensitive to the colliding system, a task which will be greatly aided by the availability of the medium interface presented in this work.

        \subsection{Small systems - a motivation}

        Decades of ultra-relativistic heavy ion collisions at colliders across the world have led to a successful program of creating and beginning to characterize the hot deconfined state of matter known as the quark-gluon plasma (QGP) \cite{Busza:2018rrf}.
        The study of the QGP falls largely into two categories:
        (1) phenomena governing low-momentum particles, and
        (2) phenomena governing high-momentum particles.
        
        It is critical that the two categories of phenomena are described by a self-consistent ``standard model of the QGP'' \cite{Gyulassy:2004vg}.
        In very central \pbpb or \auau collisions, this seems to be the case, but it has been known for several years now that the framework is inconsistent in small colliding systems such as \ppb or \dau:
        Low-momentum signatures of the QGP in small colliding systems are characterized by several observables that are both measured experimentally with a high degree of precision and described exceptionally well theoretically (see \cite{Nagle:2018nvi} for a paedegocial review);
        On the other hand, the high-momentum, or \highpt, signatures are not only absent, but their absence remains unexplained.
    
        This last statement needs to be refined: While other observables exist, the gold standard for the observation of the modification of \highpt partons by the QGP has been the nuclear modification factor, \raa\! (see \cite{Connors:2017ptx} for a review of jet measurements in heavy-ion collisions).
        \raa compares the yield in a nucleus-nucleus (\aa\!) collision, with that in  \pp and attempts to scale this ratio such that $R_{AA}\sim 1$ in the absence of the QGP.
        The observable \rpa (and related observables) does the same for a proton-nucleus collision. 
        Experimental measurements all but rule out the possibility that $R_{pA} \nsim 1$ (see \cite{Apolinario:2022vzg} and citations therein).    
        
        There can be several explanations for this inconsistency:
        It may be that the model description of \lowpt phenomena using the fluid mechanics reminiscent of the QGP in \aa are going beyond their range of applicability and should not be interpreted as evidence for collective behaviour.  This scenario has been studied extensively \cite{Heinz:2019dbd,Ambrus:2022qya}.
        The correlations may be due to non-flow effects such as momentum anisotropies that exist already in the initial state \cite{Giacalone:2020byk}.
        It seems clear that, even within hydrodynamic models that are able to describe the \lowpt phenomena, the nature of the medium in a small system must necessarily be hotter and denser in order to produce multiplicities that are comparable to peripheral \aa collisions \cite{Sievert:2019zjr}.
        
        However, the interpretation of the apparent absence of the modification of \highpt partons in small systems is not as sophisticated.
        Theoretically, it may be that the distance a hard parton travels through the medium is too short to see any modification. 
        If this were the case, then the perturbative calculations that lead to this intuition would bear it out.
        As it turns out, some studies have been done that show either that the modification should be observable \cite{Park:2016jap}, or that the array of simplifying assumptions made in the standard pQCD calculations are completely inconsistent with small systems, rendering them unreliable \cite{Kolbe:2015rvk}.
        It may be that the modification of \highpt partons occurs on a time-scale that is much larger than the lifetime of the QGP in small systems, i.e. that the mean free path is smaller than the system size.
        This is a prediction that should easily be verifiable in Monte Carlo (MC) simulations, and is partly the motivation for the present work.

        But there is a larger phenomenological obstacle to our understanding of the modification of \highpt partons in small systems:
        \raa (or its equivalent in a small system such as \rpa) is a wholly unsuitable observable.
        The bulk of the problem lies in \raa\!'s reliance on yields that are sensitive to a host of biases that are accentuated in small systems.
        These include the immense experimental difficulty of determining the number of binary collisions \cite{ALICE:2014xsp}, uncertainties in the fragmentation of jets, and the initial production spectrum through the nuclear parton distribution functions (nPDFs).
        In addition to these biases, the steeply falling production spectrum means that the necessarily small amount of energy loss rapidly becomes undetectable, even if it was present
        \!\footnote{Some alternatives exist \cite{Brewer:2021hmh,Brewer:2018dfs} that attempt to reduce the biases introduced by the steeply falling spectrum.}.
        \raa is not able to falsify the claim that a medium of deconfined QCD matter is produced in very central small systems.

        It is clear that, whatever the physics of the modification of \highpt partons in small systems is, a much more sophisticated understanding of sensitive, differential observables is needed.
        It is also crucial that such an understanding is developed in the context of experimentally achievable goals.

        \subsection{Jet substructure}
        
        Jet substructure observables suffer from far fewer of the biases that plague \raa\!.
        Jet substructure has been studied extensively in high-energy particle physics \cite{Marzani:2019hun,Larkoski:2017jix,Kogler:2018hem}, enjoying enormous success.
        There have also been many theoretical advances in the study of jet substructure in heavy-ion physics \cite{Casalderrey-Solana:2019ubu,Mulligan:2020tim,Apolinario:2017qay},
        along with several promising experamental measurements \cite{ALargeIonColliderExperiment:2021mqf,CMS:2017qlm,STAR:2021kjt}
        It is hoped that the present work will serve as a useful tool to aid the community to develop new substructure observables that characterize the modification of \highpt partons in heavy-ion collisions in a more precise manner than \raa\!.
        It is not unreasonable to presume that such precision observables will shed significant light on the modification of \highpt partons in small systems as well.

        Of particular importance in the study of jet substrcture in hadronic collisions has been the development of appropriate grooming  techniques.
        Careful consideration of grooming techniques in studies involving \jewel\, are particularly important
        \!\footnote{This statement is independent of the need to include an appropriate subtraction technique when using MC data generated by \jewel\, when keeping track of recoils. See \cite{Milhano:2022kzx} for details on the constituent subtraction method employed in this work.
        Once a \jewel\, event has been appropriately subtracted, the further use of jet grooming is phenomenological.}
        since the soft particles in a \jewel\, event record are produced by PYTHIA and are independent of the jets evolved by \jewel \,(except when including recoils in \jewel).

        The most widely used grooming technique in heavy-ions is the SoftDrop \cite{Larkoski:2014wba} technique.
        A newer grooming technique is that of Dynamical Grooming \cite{Mehtar-Tani:2019rrk}, which avoids the absolute scale cut-off employed by SoftDrop by using instead the hardest branch in a Cambridge/Aachen (C/A) re-clustering sequence to determine how to groom a jet.
        The only parameter used in Dynamical Grooming is called $a$, and sets the definition of the term ``hardest branch''.
        That is, the hardest splitting in an angular ordered shower is defined as
            \begin{equation}
                \kappa^{a} =\frac{1}{p_T}\max_{i\in \text{C/A seq.}}
                            \left[
                            z_i(1-z_i)p_{T,i}\left(\frac{\theta_i}{R}\right)^a,
                            \right]
            \end{equation}
        where $p_T$ is the transverse momentum of an entire jet with radius $R$, for $z_i$, $p_{T,i}$, and $\theta_i$ the momentum sharing fraction, the energy of the parent, and the relative splitting angle of the $i^{th}$ splitting in the C/A \cite{Dokshitzer:1997in} re-clustering sequence respectively.
        By choosing (potentially continuous) values for $a$, one varies the definition of the ``hardest branch''.  
        Through $a$, the hardest branch is defined as
        \begin{itemize}
            \item $a=0$: the branch with the most symmetric momentum sharing (use $a=0.1$ to avoid colliniear sensitivity);
            \item $a=1$: the branch with the largest relative transverse momentum;
            \item $a=2$: the branch with the shortest formation time.
        \end{itemize} 

        The dynamical grooming procedure can then be used either to tag a particular hard splitting in order to study its kinematics, or to groom the jet by discarding all emissions that occur prior to the hard splitting in the C/A sequence.
        Of course, this grooming procedure still assumes angular ordering of emissions, which is not guaranteed in a heavy-ion collision.

        In addition to the dynamically groomed jet momentum sharing fraction $z_G$, three other observables are also presented in this section: the invariant jet mass $M_{jet}$ (of a groomed jet, not the groomed jet mass), the number of subjets, and the lund plane.
        The number of subjets is obtained by first clustering $R=0.4$ jets in an event before, for each $R=0.4$ jet, reclustering the constituents into $R=0.2$ jets.
        The lund plane is computed using the FASTJET contrib package ``LundPlane'' \cite{Dreyer:2018nbf}.

        \begin{figure}
             \centering
             \begin{subfigure}[b]{0.48\textwidth}
                 \centering
                 \includegraphics[width=\textwidth]{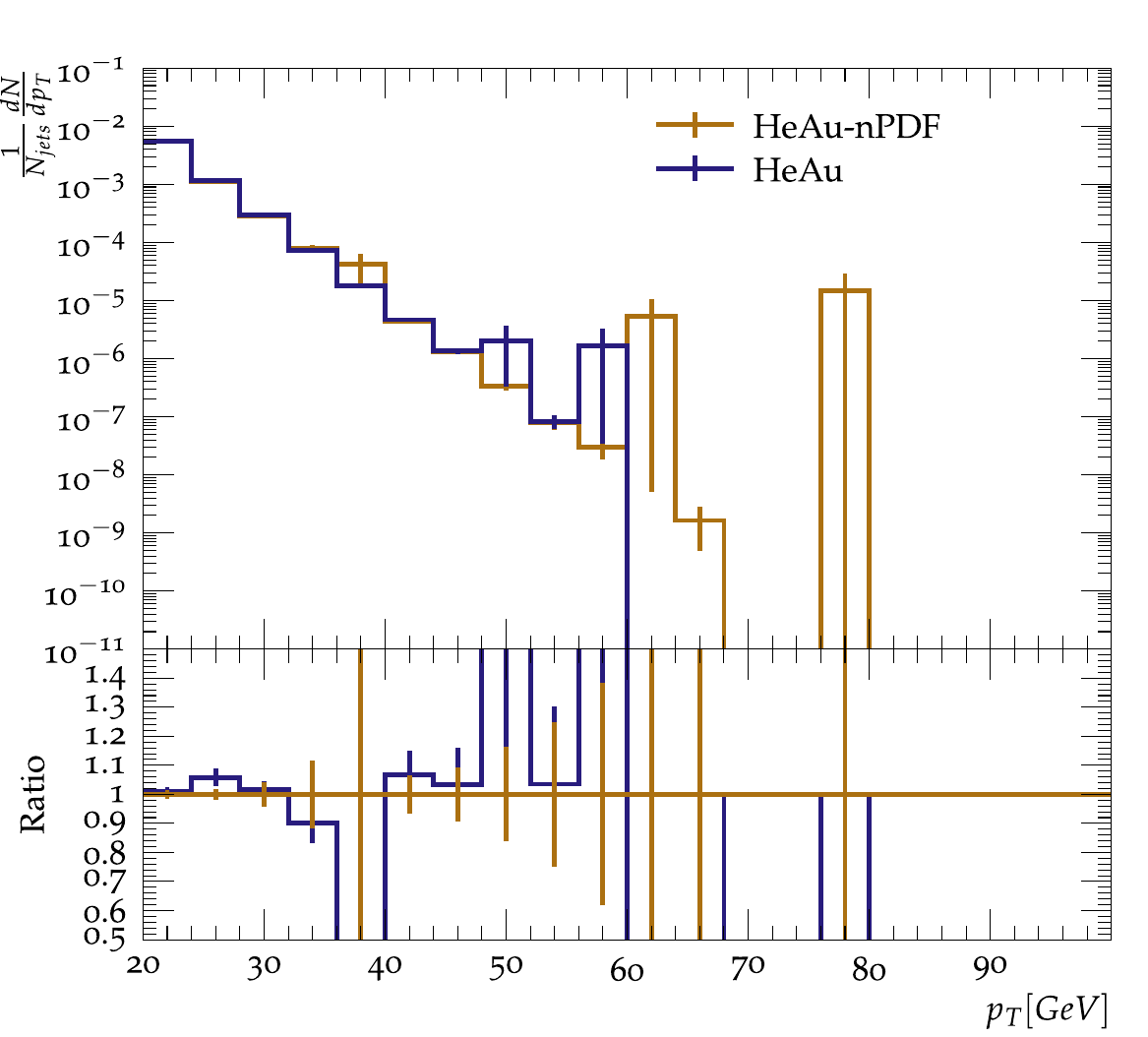}
                 \caption{\label{fig:ptDyn01}}
             \end{subfigure}
             \begin{subfigure}[b]{0.48\textwidth} 
                 \centering
                 \includegraphics[width=\textwidth]{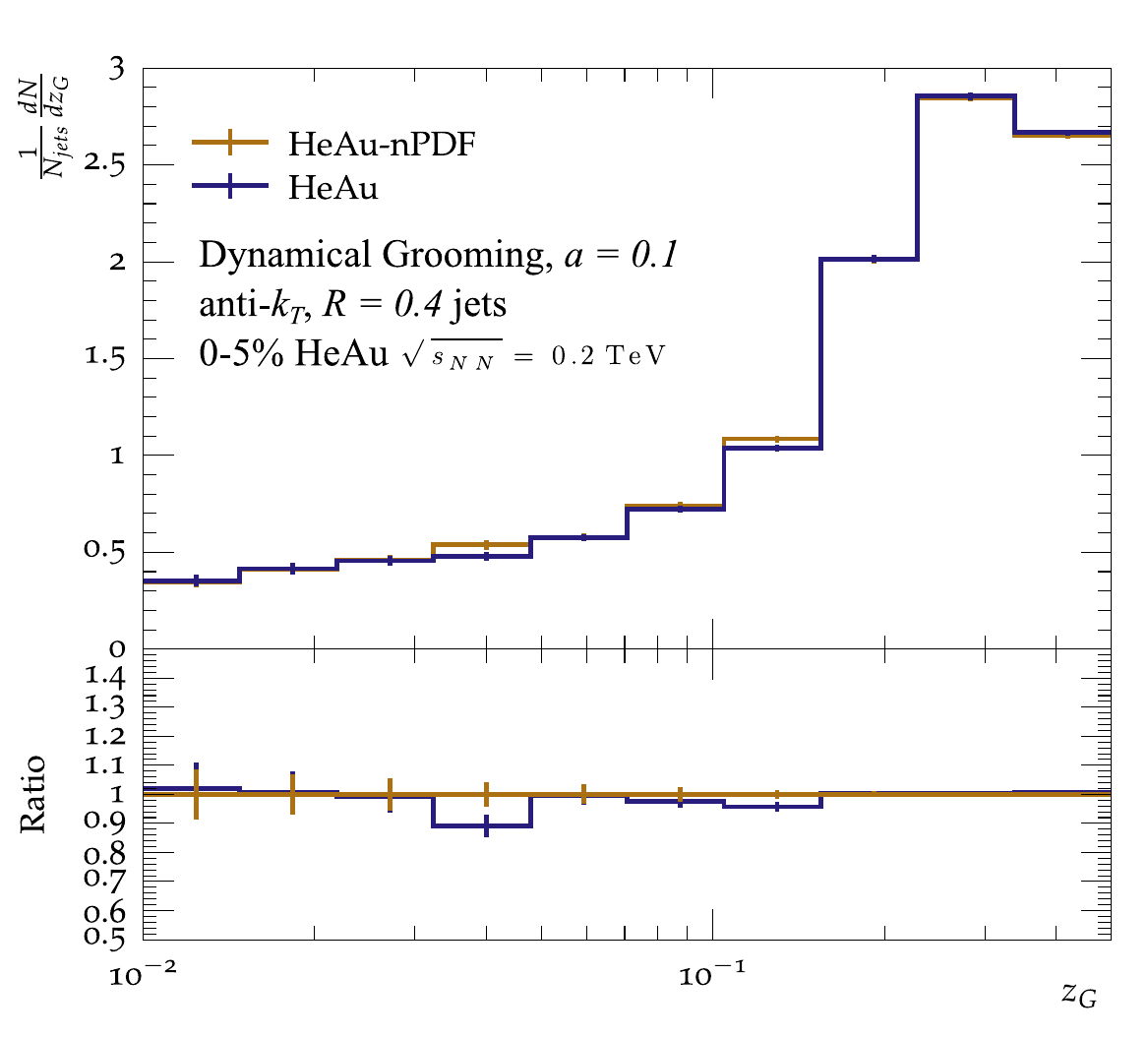}
                 \caption{\label{fig:zGDyn0}}
             \end{subfigure}
                \caption{(a) The jet yield and (b) the dynamically groomed ($a=0.1$) jet momentum sharing fraction for $R=0.4$ anti-$k_T$ jets in the $0-5\%$ most central bin of \heau collisions at $\sqrt{s_{NN}}=0.2$\,TeV. Gold histograms show simulated results using a gold nuclear PDF, while purple histograms show simulated results run on \jewel's default proton PDF setting. }
                \label{fig:PDF}
        \end{figure}
        In order to illustrate the use of the 2D medium interface presented here, simulations were performed using
        publicly available (2+1)D MUSIC \cite{Schenke:2020mbo} profiles with IP-Glasma initial conditions of the $0-5\%$ most central \heau at $\sqrt{s_{NN}}=200$GeV and \pbpb at $\sqrt{s_{NN}}=2.76$TeV collisions.
        Note that the $0-5\%$ most central \heau bin is precisely the centrality bin in which PHENIX measured the hierarchy of $v_{2,3}$ \cite{PHENIX:2015idk}. 
        For both \pbpb and \heau, event-by-event fluctuations were simulated by running 2000 events on each of 100 (200) simulation profiles for \pbpb(\heau).
        
        Consider first the effect of the choice of PDF in  \cref{fig:PDF}.  
        \Cref{fig:PDF} shows two different observables for dynamically groomed ($a=0.1$) anti-$k_T$, $R=0.4$ jets, simulated using the 2D medium interface with MUSIC profiles for \heau collisions at $\sqrt{s_{NN}}=200$\,GeV.
        Each panel shows two histograms:
        The dark purple histogram is obtained running with \jewel's default proton PDF (CT14nlo \cite{Dulat:2015mca}), while the light gold histogram was obtained running with a gold nuclear PDF (EPPS21 + CT18ANLO \cite{Eskola:2021nhw,Hou:2019efy}).
        Even for observables that are expected to be sensitive to the PDF, like the jet spectrum, the simulation agrees within Monte Carlo errors.

            \begin{figure}
             \centering
             \begin{subfigure}[b]{0.45\textwidth}
                 \centering
                 \includegraphics[width=\textwidth]{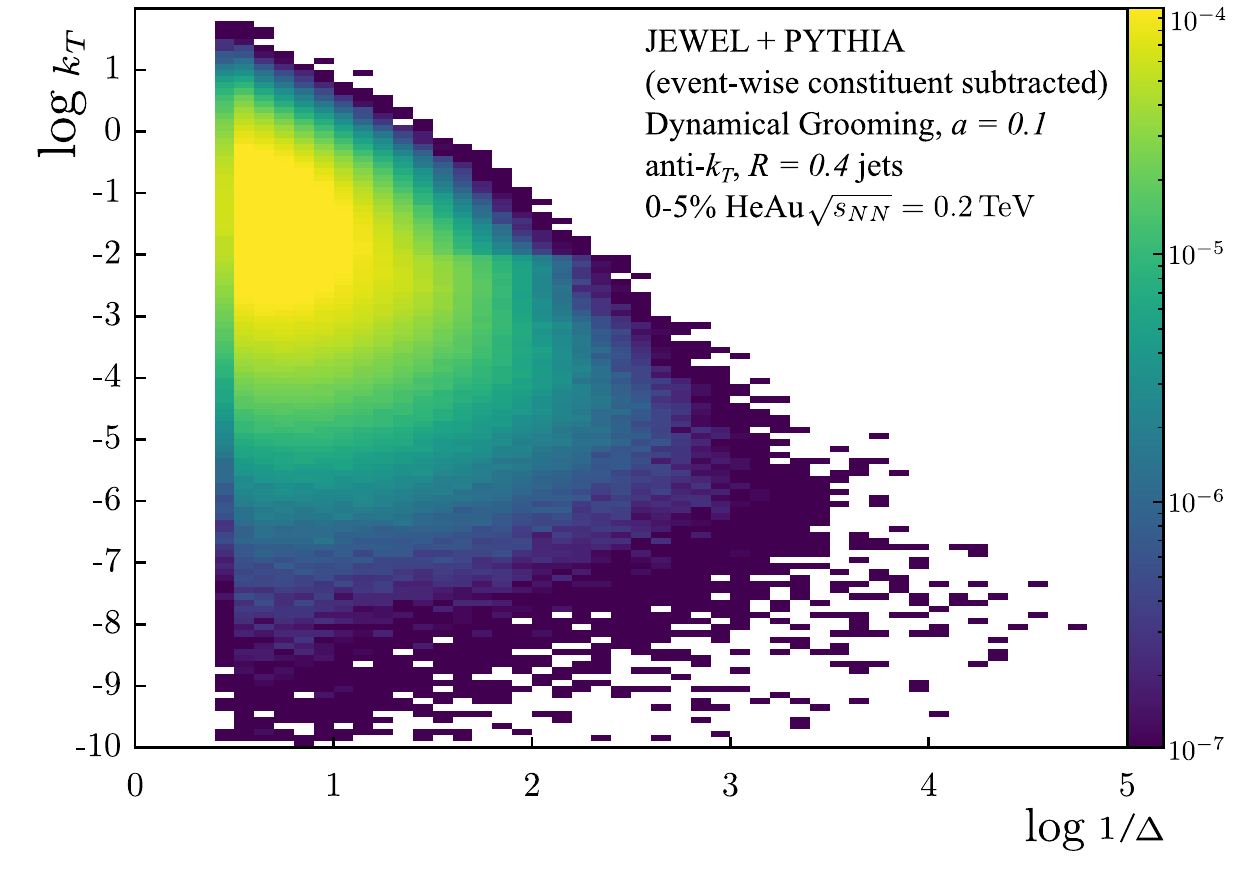}
                 \caption{\label{fig:HeAuLund}}
             \end{subfigure}
             \begin{subfigure}[b]{0.53\textwidth} 
                 \centering
                 \includegraphics[width=\textwidth]{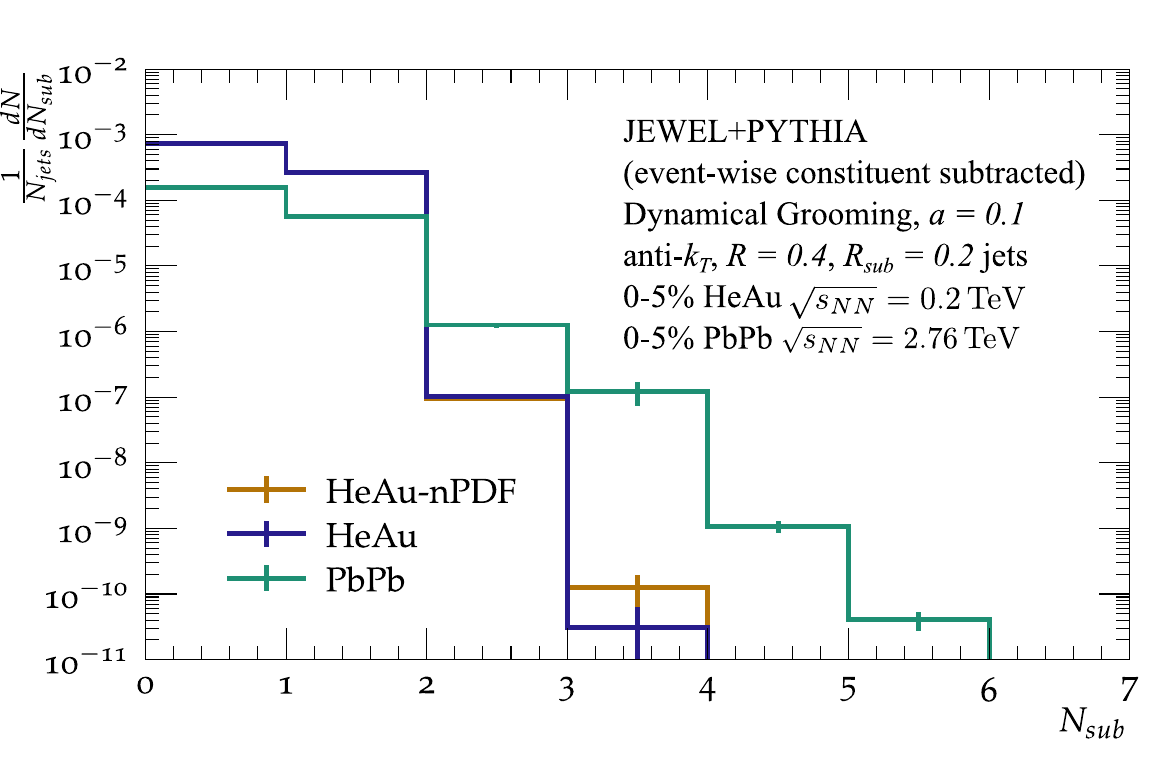}
                 \caption{\label{fig:nSub_Dyn01}}
             \end{subfigure}
                \caption{The (a) Lund Plane and (b) number of subjets  of dynamically groomed ($a=0.1$), $R=0.4$ anti-$k_T$ jets in the $0-5\%$ most central bins of \heau collisions at $\sqrt{s_{NN}}=0.2$\,TeV and \pbpb collisions at $\sqrt{s_{NN}}=2.76$\,TeV. Gold histograms show simulated results for \heau using a gold nuclear PDF, purple histograms for \heau using \jewel's default proton PDF setting, and blue histograms for \pbpb. }
                \label{fig:substructure}
        \end{figure}

        As an illustrative example of standard substructure observables, consider \cref{fig:substructure}, showing the Lund Plane in the left panel and the number of subjets with radius $R=0.2$ in the right panel, for dynamically groomed ($a=0.1$), anti-$k_T$ jets with radius of $R=0.4$. 
        Dark purple histograms are obtained running with \jewel's default proton PDF (CT14nlo \cite{Dulat:2015mca}), light gold histograms were obtained running with a gold nuclear PDF (EPPS21 + CT18ANLO \cite{Eskola:2021nhw,Hou:2019efy}), on MUSIC \heau profiles, blue histograms were obtained running \jewel\, (default proton PDF) on MUSIC \pbpb profiles.

            \begin{figure}
                \centering
                \includegraphics[scale=0.17]{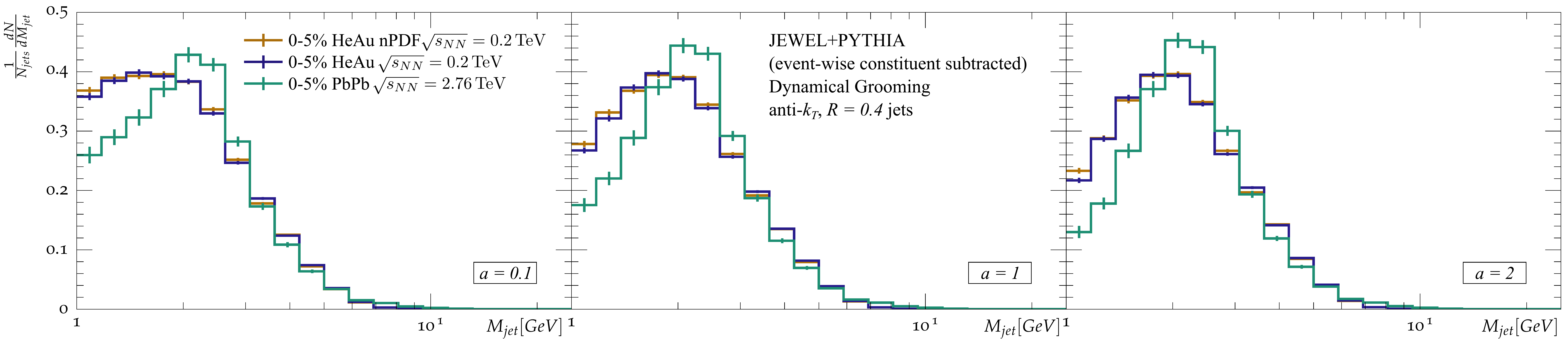}
                \caption{The jet mass for dynamically groomed, $R=0.4$ anti-$k_T$ jets in the $0-5\%$ most central bins of \heau collisions at $\sqrt{s_{NN}}=0.2$\,TeV and \pbpb collisions at $\sqrt{s_{NN}}=2.76$\,TeV. Gold histograms show simulated results for \heau using a gold nuclear PDF, purple histograms for \heau using \jewel's default proton PDF setting, and blue histograms for \pbpb. From left to right the panels use a grooming parameter of $a=0.1$, $a=1$, and $a=2$ respectively, corresponding to different definitions of the ``hardest branch''.}
                \label{fig:jetMass_Dyn_all}
            \end{figure}

        The constituent subtraction method has been validated using the jet mass, improving the interpretability of the shape of the jet mass distribution \cite{Milhano:2022kzx}.
        Consider therefore the variation in the shape due to the choice of grooming parameter $a$ in \cref{fig:jetMass_Dyn_all}.

            \begin{figure}
                \centering
                \includegraphics[scale=0.17]{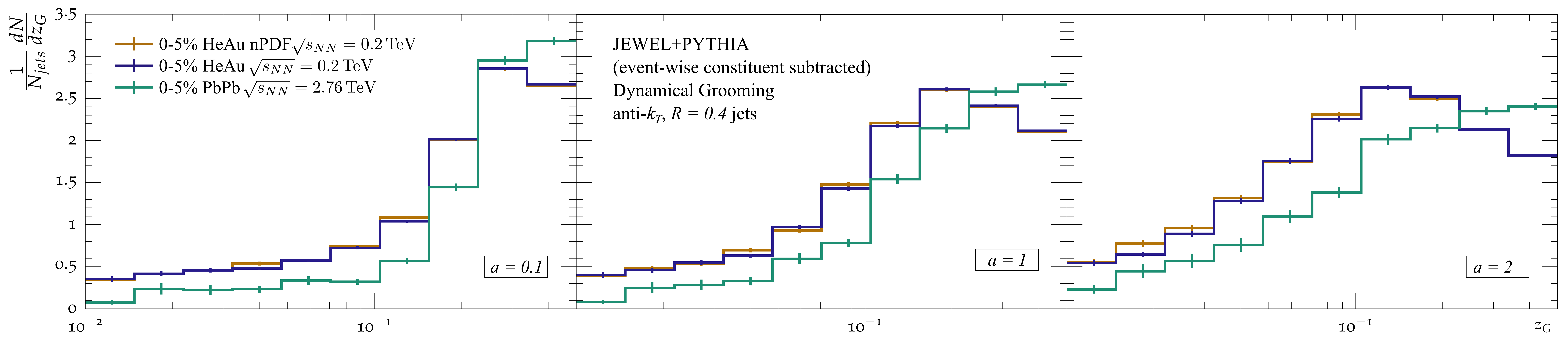}
                \caption{The groomed momentum sharing fraction for dynamically groomed, $R=0.4$ anti-$k_T$ jets in the $0-5\%$ most central bins of \heau collisions at $\sqrt{s_{NN}}=0.2$\,TeV and \pbpb collisions at $\sqrt{s_{NN}}=2.76$\,TeV. Gold histograms show simulated results for \heau using a gold nuclear PDF, purple histograms for \heau using \jewel's default proton PDF setting, and blue histograms for \pbpb. From left to right the panels use a grooming parameter of $a=0.1$, $a=1$, and $a=2$ respectively, corresponding to different definitions of the ``hardest branch''.}
                \label{fig:zG_Dyn_all}
            \end{figure}
            
        Lastly, the groomed momentum sharing fraction is shown in \cref{fig:zG_Dyn_all} for dynamically groomed anti-$k_T$ jets with radius of $R=0.4$, for the three standard choices of $a$.
        It is curious to note the peak in the \heau samples in the $a=1$ and $a=2$ panels of \cref{fig:zG_Dyn_all} which is absent for the \pbpb samples.
        An exploration of the origins of this feature is left for later work as it will involve the higher statistics needed to perform studies that are differential in the jet energy.

    \section{Conclusion and Outlook}

        The purpose of this letter is to present a medium interface for \jewel\, that allows \jewel\, to sample a given set of temperature and velocity profiles of a (2+1)D medium.
        The interface is made publicly available.
        It is hoped that this tool will be of use to the broader heavy-ion jet community to aid in the exploration of jets in a variety of collision systems.

        There is an important aspect of heavy-ion collisions that is not yet taken into account by this interface - the underlying event.
        In its current form, this medium interface relies, in precisely the same way as the standard \jewel\, medium model, on a completely uncorrelated underlying event produced by PYTHIA.
        Although one may gain some access to the part of the event that is correlated with the jet by keeping track of recoiling partons, it would still be desirable to have, in the \jewel\, event record, the underlying event that is the result of the medium upon which the jet was evolved.
        It is worth noting that, without the information from the underlying event, it is not possible to study any observable that relies on the soft constituents of the event, such as \highpt-$v_2$.
        Such a modification of the main \jewel\, code is a much larger undertaking and is left for future work.

        Although much focus has been places here on the use of this interface to study small systems, its usefulness extends to any precision study of the jet-medium interaction.
        There is particular scope to vary the nature of the medium using this interface, which allows for a cleanly interpretable exploration of aspects of jet evolution that are sensitive to the nature of the medium, not simply the existence thereof.

    \section{Acknowledgments}

        I gratefully acknowledge the major contribution to the original implementation of the medium interface by Korinna Zapp.
        Thank you to Alba Soto-Ontoso for providing a C++ class that implements the dynamical grooming procedure, and to Liliana Apolin\'{a}rio for help implementing the constituent subtraction.
        Thank you to Chun Shen for support in accessing the MUSIC profiles. 
        
        I would like to express gratitude to Urs Wiedemann, Guilherme Milhano and Korinna Zapp for early discussions that lead to this project being started, as well as to the CERN TH department for their hospitality during the start of this project, and the SA-CERN collaboration for funding to visit CERN.
        
        I am grateful for many fruitful discussions with, and helpful suggestions from, Leonardo Barreto, Fabio Canedo and Marcelo Munhoz at the University of Sao Paulo.
        This project has further benefited greatly from insights garnered through conversations with 
        Liliana Apolinário,
        Jasmine Brewer,
        Raghav Kunnawalkam Elayavalli,
        Mawande Lushozi, 
        Anna McCoy,
        Titus Momb\"{a}cher,
        Christine Nattrass,
        Dmytro Oliinychenko, and
        Carlos Salgado.
        
        Lastly, the development of this project relied heavily on access to resources at both the Institute for Nuclear Theory at the University of Washington (and was therefore partially supported by the U.S. DOE under Grant No. DE-FG02-00ER41132) and the Galician Institute for High Energy Physics at the University of Santiago de Compostela; I would like to express particular gratitude to the general and computing administrative staff at both institutions.

         This work is supported by European Research Council project ERC-2018-ADG-835105 YoctoLHC; by Maria de Maetzu excellence program under project CEX2020-001035-M; by Spanish Research State Agency under project PID2020-119632GB- I00; and by Xunta de Galicia (Centro singular de investigacion de Galicia accreditation 2019-2022), by European Union ERDF.)

    
    \bibliography{main}

\begin{thebibliography}{45}%
\makeatletter
\providecommand \@ifxundefined [1]{%
 \@ifx{#1\undefined}
}%
\providecommand \@ifnum [1]{%
 \ifnum #1\expandafter \@firstoftwo
 \else \expandafter \@secondoftwo
 \fi
}%
\providecommand \@ifx [1]{%
 \ifx #1\expandafter \@firstoftwo
 \else \expandafter \@secondoftwo
 \fi
}%
\providecommand \natexlab [1]{#1}%
\providecommand \enquote  [1]{``#1''}%
\providecommand \bibnamefont  [1]{#1}%
\providecommand \bibfnamefont [1]{#1}%
\providecommand \citenamefont [1]{#1}%
\providecommand \href@noop [0]{\@secondoftwo}%
\providecommand \href [0]{\begingroup \@sanitize@url \@href}%
\providecommand \@href[1]{\@@startlink{#1}\@@href}%
\providecommand \@@href[1]{\endgroup#1\@@endlink}%
\providecommand \@sanitize@url [0]{\catcode `\\12\catcode `\$12\catcode
  `\&12\catcode `\#12\catcode `\^12\catcode `\_12\catcode `\%12\relax}%
\providecommand \@@startlink[1]{}%
\providecommand \@@endlink[0]{}%
\providecommand \url  [0]{\begingroup\@sanitize@url \@url }%
\providecommand \@url [1]{\endgroup\@href {#1}{\urlprefix }}%
\providecommand \urlprefix  [0]{URL }%
\providecommand \Eprint [0]{\href }%
\providecommand \doibase [0]{https://doi.org/}%
\providecommand \selectlanguage [0]{\@gobble}%
\providecommand \bibinfo  [0]{\@secondoftwo}%
\providecommand \bibfield  [0]{\@secondoftwo}%
\providecommand \translation [1]{[#1]}%
\providecommand \BibitemOpen [0]{}%
\providecommand \bibitemStop [0]{}%
\providecommand \bibitemNoStop [0]{.\EOS\space}%
\providecommand \EOS [0]{\spacefactor3000\relax}%
\providecommand \BibitemShut  [1]{\csname bibitem#1\endcsname}%
\let\auto@bib@innerbib\@empty
\bibitem [{\citenamefont {Apolin\'ario}\ \emph {et~al.}(2022)\citenamefont
  {Apolin\'ario}, \citenamefont {Lee},\ and\ \citenamefont
  {Winn}}]{Apolinario:2022vzg}%
  \BibitemOpen
  \bibfield  {author} {\bibinfo {author} {\bibfnamefont {L.}~\bibnamefont
  {Apolin\'ario}}, \bibinfo {author} {\bibfnamefont {Y.-J.}\ \bibnamefont
  {Lee}},\ and\ \bibinfo {author} {\bibfnamefont {M.}~\bibnamefont {Winn}},\
  }\bibfield  {title} {\bibinfo {title} {{Heavy quarks and jets as probes of
  the QGP}},\ }\href {https://doi.org/10.1016/j.ppnp.2022.103990} {\bibfield
  {journal} {\bibinfo  {journal} {Prog. Part. Nucl. Phys.}\ }\textbf {\bibinfo
  {volume} {127}},\ \bibinfo {pages} {103990} (\bibinfo {year} {2022})},\
  \Eprint {https://arxiv.org/abs/2203.16352} {arXiv:2203.16352 [hep-ph]}
  \BibitemShut {NoStop}%
\bibitem [{\citenamefont {Zapp}\ \emph {et~al.}(2013)\citenamefont {Zapp},
  \citenamefont {Krauss},\ and\ \citenamefont {Wiedemann}}]{Zapp:2012ak}%
  \BibitemOpen
  \bibfield  {author} {\bibinfo {author} {\bibfnamefont {K.~C.}\ \bibnamefont
  {Zapp}}, \bibinfo {author} {\bibfnamefont {F.}~\bibnamefont {Krauss}},\ and\
  \bibinfo {author} {\bibfnamefont {U.~A.}\ \bibnamefont {Wiedemann}},\
  }\bibfield  {title} {\bibinfo {title} {{A perturbative framework for jet
  quenching}},\ }\href {https://doi.org/10.1007/JHEP03(2013)080} {\bibfield
  {journal} {\bibinfo  {journal} {JHEP}\ }\textbf {\bibinfo {volume} {03}},\
  \bibinfo {pages} {080}},\ \Eprint {https://arxiv.org/abs/1212.1599}
  {arXiv:1212.1599 [hep-ph]} \BibitemShut {NoStop}%
\bibitem [{\citenamefont {Zapp}(2014)}]{Zapp:2013vla}%
  \BibitemOpen
  \bibfield  {author} {\bibinfo {author} {\bibfnamefont {K.~C.}\ \bibnamefont
  {Zapp}},\ }\bibfield  {title} {\bibinfo {title} {{JEWEL 2.0.0: directions for
  use}},\ }\href {https://doi.org/10.1140/epjc/s10052-014-2762-1} {\bibfield
  {journal} {\bibinfo  {journal} {Eur. Phys. J. C}\ }\textbf {\bibinfo {volume}
  {74}},\ \bibinfo {pages} {2762} (\bibinfo {year} {2014})},\ \Eprint
  {https://arxiv.org/abs/1311.0048} {arXiv:1311.0048 [hep-ph]} \BibitemShut
  {NoStop}%
\bibitem [{\citenamefont {Zapp}\ \emph {et~al.}(2009)\citenamefont {Zapp},
  \citenamefont {Ingelman}, \citenamefont {Rathsman}, \citenamefont {Stachel},\
  and\ \citenamefont {Wiedemann}}]{Zapp:2008gi}%
  \BibitemOpen
  \bibfield  {author} {\bibinfo {author} {\bibfnamefont {K.}~\bibnamefont
  {Zapp}}, \bibinfo {author} {\bibfnamefont {G.}~\bibnamefont {Ingelman}},
  \bibinfo {author} {\bibfnamefont {J.}~\bibnamefont {Rathsman}}, \bibinfo
  {author} {\bibfnamefont {J.}~\bibnamefont {Stachel}},\ and\ \bibinfo {author}
  {\bibfnamefont {U.~A.}\ \bibnamefont {Wiedemann}},\ }\bibfield  {title}
  {\bibinfo {title} {{A Monte Carlo Model for 'Jet Quenching'}},\ }\href
  {https://doi.org/10.1140/epjc/s10052-009-0941-2} {\bibfield  {journal}
  {\bibinfo  {journal} {Eur. Phys. J. C}\ }\textbf {\bibinfo {volume} {60}},\
  \bibinfo {pages} {617} (\bibinfo {year} {2009})},\ \Eprint
  {https://arxiv.org/abs/0804.3568} {arXiv:0804.3568 [hep-ph]} \BibitemShut
  {NoStop}%
\bibitem [{\citenamefont {Bierlich}\ \emph {et~al.}(2020)\citenamefont
  {Bierlich} \emph {et~al.}}]{Bierlich:2019rhm}%
  \BibitemOpen
  \bibfield  {author} {\bibinfo {author} {\bibfnamefont {C.}~\bibnamefont
  {Bierlich}} \emph {et~al.},\ }\bibfield  {title} {\bibinfo {title} {{Robust
  Independent Validation of Experiment and Theory: Rivet version 3}},\ }\href
  {https://doi.org/10.21468/SciPostPhys.8.2.026} {\bibfield  {journal}
  {\bibinfo  {journal} {SciPost Phys.}\ }\textbf {\bibinfo {volume} {8}},\
  \bibinfo {pages} {026} (\bibinfo {year} {2020})},\ \Eprint
  {https://arxiv.org/abs/1912.05451} {arXiv:1912.05451 [hep-ph]} \BibitemShut
  {NoStop}%
\bibitem [{\citenamefont {Cacciari}\ and\ \citenamefont
  {Salam}(2006)}]{Cacciari:2005hq}%
  \BibitemOpen
  \bibfield  {author} {\bibinfo {author} {\bibfnamefont {M.}~\bibnamefont
  {Cacciari}}\ and\ \bibinfo {author} {\bibfnamefont {G.~P.}\ \bibnamefont
  {Salam}},\ }\bibfield  {title} {\bibinfo {title} {{Dispelling the $N^{3}$
  myth for the $k_t$ jet-finder}},\ }\href
  {https://doi.org/10.1016/j.physletb.2006.08.037} {\bibfield  {journal}
  {\bibinfo  {journal} {Phys. Lett. B}\ }\textbf {\bibinfo {volume} {641}},\
  \bibinfo {pages} {57} (\bibinfo {year} {2006})},\ \Eprint
  {https://arxiv.org/abs/hep-ph/0512210} {arXiv:hep-ph/0512210} \BibitemShut
  {NoStop}%
\bibitem [{\citenamefont {Cacciari}\ \emph {et~al.}(2012)\citenamefont
  {Cacciari}, \citenamefont {Salam},\ and\ \citenamefont
  {Soyez}}]{Cacciari:2011ma}%
  \BibitemOpen
  \bibfield  {author} {\bibinfo {author} {\bibfnamefont {M.}~\bibnamefont
  {Cacciari}}, \bibinfo {author} {\bibfnamefont {G.~P.}\ \bibnamefont
  {Salam}},\ and\ \bibinfo {author} {\bibfnamefont {G.}~\bibnamefont {Soyez}},\
  }\bibfield  {title} {\bibinfo {title} {{FastJet User Manual}},\ }\href
  {https://doi.org/10.1140/epjc/s10052-012-1896-2} {\bibfield  {journal}
  {\bibinfo  {journal} {Eur. Phys. J. C}\ }\textbf {\bibinfo {volume} {72}},\
  \bibinfo {pages} {1896} (\bibinfo {year} {2012})},\ \Eprint
  {https://arxiv.org/abs/1111.6097} {arXiv:1111.6097 [hep-ph]} \BibitemShut
  {NoStop}%
\bibitem [{\citenamefont {Dreyer}\ \emph {et~al.}(2018)\citenamefont {Dreyer},
  \citenamefont {Salam},\ and\ \citenamefont {Soyez}}]{Dreyer:2018nbf}%
  \BibitemOpen
  \bibfield  {author} {\bibinfo {author} {\bibfnamefont {F.~A.}\ \bibnamefont
  {Dreyer}}, \bibinfo {author} {\bibfnamefont {G.~P.}\ \bibnamefont {Salam}},\
  and\ \bibinfo {author} {\bibfnamefont {G.}~\bibnamefont {Soyez}},\ }\bibfield
   {title} {\bibinfo {title} {{The Lund Jet Plane}},\ }\href
  {https://doi.org/10.1007/JHEP12(2018)064} {\bibfield  {journal} {\bibinfo
  {journal} {JHEP}\ }\textbf {\bibinfo {volume} {12}},\ \bibinfo {pages}
  {064}},\ \Eprint {https://arxiv.org/abs/1807.04758} {arXiv:1807.04758
  [hep-ph]} \BibitemShut {NoStop}%
\bibitem [{\citenamefont {Milhano}\ and\ \citenamefont
  {Zapp}(2022)}]{Milhano:2022kzx}%
  \BibitemOpen
  \bibfield  {author} {\bibinfo {author} {\bibfnamefont {J.~G.}\ \bibnamefont
  {Milhano}}\ and\ \bibinfo {author} {\bibfnamefont {K.}~\bibnamefont {Zapp}},\
  }\bibfield  {title} {\bibinfo {title} {{Improved background subtraction and a
  fresh look at jet sub-structure in JEWEL}},\ }\href
  {https://doi.org/10.1140/epjc/s10052-022-10954-1} {\bibfield  {journal}
  {\bibinfo  {journal} {Eur. Phys. J. C}\ }\textbf {\bibinfo {volume} {82}},\
  \bibinfo {pages} {1010} (\bibinfo {year} {2022})},\ \Eprint
  {https://arxiv.org/abs/2207.14814} {arXiv:2207.14814 [hep-ph]} \BibitemShut
  {NoStop}%
\bibitem [{\citenamefont {Buckley}\ \emph {et~al.}(2015)\citenamefont
  {Buckley}, \citenamefont {Ferrando}, \citenamefont {Lloyd}, \citenamefont
  {Nordstr\"om}, \citenamefont {Page}, \citenamefont {R\"ufenacht},
  \citenamefont {Sch\"onherr},\ and\ \citenamefont {Watt}}]{Buckley:2014ana}%
  \BibitemOpen
  \bibfield  {author} {\bibinfo {author} {\bibfnamefont {A.}~\bibnamefont
  {Buckley}}, \bibinfo {author} {\bibfnamefont {J.}~\bibnamefont {Ferrando}},
  \bibinfo {author} {\bibfnamefont {S.}~\bibnamefont {Lloyd}}, \bibinfo
  {author} {\bibfnamefont {K.}~\bibnamefont {Nordstr\"om}}, \bibinfo {author}
  {\bibfnamefont {B.}~\bibnamefont {Page}}, \bibinfo {author} {\bibfnamefont
  {M.}~\bibnamefont {R\"ufenacht}}, \bibinfo {author} {\bibfnamefont
  {M.}~\bibnamefont {Sch\"onherr}},\ and\ \bibinfo {author} {\bibfnamefont
  {G.}~\bibnamefont {Watt}},\ }\bibfield  {title} {\bibinfo {title} {{LHAPDF6:
  parton density access in the LHC precision era}},\ }\href
  {https://doi.org/10.1140/epjc/s10052-015-3318-8} {\bibfield  {journal}
  {\bibinfo  {journal} {Eur. Phys. J. C}\ }\textbf {\bibinfo {volume} {75}},\
  \bibinfo {pages} {132} (\bibinfo {year} {2015})},\ \Eprint
  {https://arxiv.org/abs/1412.7420} {arXiv:1412.7420 [hep-ph]} \BibitemShut
  {NoStop}%
\bibitem [{\citenamefont {Floerchinger}\ and\ \citenamefont
  {Zapp}(2014)}]{Floerchinger:2014yqa}%
  \BibitemOpen
  \bibfield  {author} {\bibinfo {author} {\bibfnamefont {S.}~\bibnamefont
  {Floerchinger}}\ and\ \bibinfo {author} {\bibfnamefont {K.~C.}\ \bibnamefont
  {Zapp}},\ }\bibfield  {title} {\bibinfo {title} {{Hydrodynamics and Jets in
  Dialogue}},\ }\href {https://doi.org/10.1140/epjc/s10052-014-3189-4}
  {\bibfield  {journal} {\bibinfo  {journal} {Eur. Phys. J. C}\ }\textbf
  {\bibinfo {volume} {74}},\ \bibinfo {pages} {3189} (\bibinfo {year}
  {2014})},\ \Eprint {https://arxiv.org/abs/1407.1782} {arXiv:1407.1782
  [hep-ph]} \BibitemShut {NoStop}%
\bibitem [{\citenamefont {Zapp}\ and\ \citenamefont
  {Floerchinger}(2014)}]{Zapp:2014msa}%
  \BibitemOpen
  \bibfield  {author} {\bibinfo {author} {\bibfnamefont {K.~C.}\ \bibnamefont
  {Zapp}}\ and\ \bibinfo {author} {\bibfnamefont {S.}~\bibnamefont
  {Floerchinger}},\ }\bibfield  {title} {\bibinfo {title} {{Interplay between
  hydrodynamics and jets}},\ }\href
  {https://doi.org/10.1016/j.nuclphysa.2014.09.037} {\bibfield  {journal}
  {\bibinfo  {journal} {Nucl. Phys. A}\ }\textbf {\bibinfo {volume} {931}},\
  \bibinfo {pages} {388} (\bibinfo {year} {2014})},\ \Eprint
  {https://arxiv.org/abs/1408.0903} {arXiv:1408.0903 [hep-ph]} \BibitemShut
  {NoStop}%
\bibitem [{\citenamefont {Barreto}\ \emph {et~al.}(2022)\citenamefont
  {Barreto}, \citenamefont {Canedo}, \citenamefont {Munhoz}, \citenamefont
  {Noronha},\ and\ \citenamefont {Noronha-Hostler}}]{Barreto:2022ulg}%
  \BibitemOpen
  \bibfield  {author} {\bibinfo {author} {\bibfnamefont {L.}~\bibnamefont
  {Barreto}}, \bibinfo {author} {\bibfnamefont {F.~M.}\ \bibnamefont {Canedo}},
  \bibinfo {author} {\bibfnamefont {M.~G.}\ \bibnamefont {Munhoz}}, \bibinfo
  {author} {\bibfnamefont {J.}~\bibnamefont {Noronha}},\ and\ \bibinfo {author}
  {\bibfnamefont {J.}~\bibnamefont {Noronha-Hostler}},\ }\bibfield  {title}
  {\bibinfo {title} {{Jet cone radius dependence of $R_{AA}$ and $v_2$ at PbPb
  5.02 TeV from JEWEL+$\rm T_RENTo$+v-USPhydro}},\ }\href@noop {} {\  (\bibinfo
  {year} {2022})},\ \Eprint {https://arxiv.org/abs/2208.02061}
  {arXiv:2208.02061 [nucl-th]} \BibitemShut {NoStop}%
\bibitem [{\citenamefont {Canedo}(2020)}]{Canedo:2020xzf}%
  \BibitemOpen
  \bibfield  {author} {\bibinfo {author} {\bibfnamefont {F.}~\bibnamefont
  {Canedo}},\ }\bibfield  {title} {\bibinfo {title} {{Study of Jet Quenching in
  Relativistic Heavy-Ion Collisions}},\ }\href@noop {} {\  (\bibinfo {year}
  {2020})},\ \Eprint {https://arxiv.org/abs/2005.13010} {arXiv:2005.13010
  [hep-ph]} \BibitemShut {NoStop}%
\bibitem [{\citenamefont {Barreto}(2021)}]{Barreto:2021fbt}%
  \BibitemOpen
  \bibfield  {author} {\bibinfo {author} {\bibfnamefont {L.}~\bibnamefont
  {Barreto}},\ }\emph {\bibinfo {title} {{Study of Jet Modification in
  Relativistic Heavy-Ion Collisions}}},\ \href
  {https://doi.org/10.11606/D.43.2021.tde-05112021-191914} {Master's thesis},\
  \bibinfo  {school} {Sao Paulo U.} (\bibinfo {year} {2021})\BibitemShut
  {NoStop}%
\bibitem [{\citenamefont {Busza}\ \emph {et~al.}(2018)\citenamefont {Busza},
  \citenamefont {Rajagopal},\ and\ \citenamefont {van~der
  Schee}}]{Busza:2018rrf}%
  \BibitemOpen
  \bibfield  {author} {\bibinfo {author} {\bibfnamefont {W.}~\bibnamefont
  {Busza}}, \bibinfo {author} {\bibfnamefont {K.}~\bibnamefont {Rajagopal}},\
  and\ \bibinfo {author} {\bibfnamefont {W.}~\bibnamefont {van~der Schee}},\
  }\bibfield  {title} {\bibinfo {title} {{Heavy Ion Collisions: The Big
  Picture, and the Big Questions}},\ }\href
  {https://doi.org/10.1146/annurev-nucl-101917-020852} {\bibfield  {journal}
  {\bibinfo  {journal} {Ann. Rev. Nucl. Part. Sci.}\ }\textbf {\bibinfo
  {volume} {68}},\ \bibinfo {pages} {339} (\bibinfo {year} {2018})},\ \Eprint
  {https://arxiv.org/abs/1802.04801} {arXiv:1802.04801 [hep-ph]} \BibitemShut
  {NoStop}%
\bibitem [{\citenamefont {Gyulassy}(2004)}]{Gyulassy:2004vg}%
  \BibitemOpen
  \bibfield  {author} {\bibinfo {author} {\bibfnamefont {M.}~\bibnamefont
  {Gyulassy}},\ }\bibfield  {title} {\bibinfo {title} {{The QGP discovered at
  RHIC}},\ }in\ \href@noop {} {\emph {\bibinfo {booktitle} {{NATO Advanced
  Study Institute: Structure and Dynamics of Elementary Matter}}}}\ (\bibinfo
  {year} {2004})\ pp.\ \bibinfo {pages} {159--182},\ \Eprint
  {https://arxiv.org/abs/nucl-th/0403032} {arXiv:nucl-th/0403032} \BibitemShut
  {NoStop}%
\bibitem [{\citenamefont {Nagle}\ and\ \citenamefont
  {Zajc}(2018)}]{Nagle:2018nvi}%
  \BibitemOpen
  \bibfield  {author} {\bibinfo {author} {\bibfnamefont {J.~L.}\ \bibnamefont
  {Nagle}}\ and\ \bibinfo {author} {\bibfnamefont {W.~A.}\ \bibnamefont
  {Zajc}},\ }\bibfield  {title} {\bibinfo {title} {{Small System Collectivity
  in Relativistic Hadronic and Nuclear Collisions}},\ }\href
  {https://doi.org/10.1146/annurev-nucl-101916-123209} {\bibfield  {journal}
  {\bibinfo  {journal} {Ann. Rev. Nucl. Part. Sci.}\ }\textbf {\bibinfo
  {volume} {68}},\ \bibinfo {pages} {211} (\bibinfo {year} {2018})},\ \Eprint
  {https://arxiv.org/abs/1801.03477} {arXiv:1801.03477 [nucl-ex]} \BibitemShut
  {NoStop}%
\bibitem [{\citenamefont {Connors}\ \emph {et~al.}(2018)\citenamefont
  {Connors}, \citenamefont {Nattrass}, \citenamefont {Reed},\ and\
  \citenamefont {Salur}}]{Connors:2017ptx}%
  \BibitemOpen
  \bibfield  {author} {\bibinfo {author} {\bibfnamefont {M.}~\bibnamefont
  {Connors}}, \bibinfo {author} {\bibfnamefont {C.}~\bibnamefont {Nattrass}},
  \bibinfo {author} {\bibfnamefont {R.}~\bibnamefont {Reed}},\ and\ \bibinfo
  {author} {\bibfnamefont {S.}~\bibnamefont {Salur}},\ }\bibfield  {title}
  {\bibinfo {title} {{Jet measurements in heavy ion physics}},\ }\href
  {https://doi.org/10.1103/RevModPhys.90.025005} {\bibfield  {journal}
  {\bibinfo  {journal} {Rev. Mod. Phys.}\ }\textbf {\bibinfo {volume} {90}},\
  \bibinfo {pages} {025005} (\bibinfo {year} {2018})},\ \Eprint
  {https://arxiv.org/abs/1705.01974} {arXiv:1705.01974 [nucl-ex]} \BibitemShut
  {NoStop}%
\bibitem [{\citenamefont {Heinz}\ and\ \citenamefont
  {Moreland}(2019)}]{Heinz:2019dbd}%
  \BibitemOpen
  \bibfield  {author} {\bibinfo {author} {\bibfnamefont {U.~W.}\ \bibnamefont
  {Heinz}}\ and\ \bibinfo {author} {\bibfnamefont {J.~S.}\ \bibnamefont
  {Moreland}},\ }\bibfield  {title} {\bibinfo {title} {{Hydrodynamic flow in
  small systems or: \textquotedblleft{}How the heck is it possible that a
  system emitting only a dozen particles can be described by fluid
  dynamics?\textquotedblright{}}},\ }\href
  {https://doi.org/10.1088/1742-6596/1271/1/012018} {\bibfield  {journal}
  {\bibinfo  {journal} {J. Phys. Conf. Ser.}\ }\textbf {\bibinfo {volume}
  {1271}},\ \bibinfo {pages} {012018} (\bibinfo {year} {2019})},\ \Eprint
  {https://arxiv.org/abs/1904.06592} {arXiv:1904.06592 [nucl-th]} \BibitemShut
  {NoStop}%
\bibitem [{\citenamefont {Ambrus}\ \emph {et~al.}(2022)\citenamefont {Ambrus},
  \citenamefont {Schlichting},\ and\ \citenamefont
  {Werthmann}}]{Ambrus:2022qya}%
  \BibitemOpen
  \bibfield  {author} {\bibinfo {author} {\bibfnamefont {V.~E.}\ \bibnamefont
  {Ambrus}}, \bibinfo {author} {\bibfnamefont {S.}~\bibnamefont
  {Schlichting}},\ and\ \bibinfo {author} {\bibfnamefont {C.}~\bibnamefont
  {Werthmann}},\ }\bibfield  {title} {\bibinfo {title} {{Establishing the range
  of applicability of hydrodynamics in high-energy collisions}},\ }\href@noop
  {} {\  (\bibinfo {year} {2022})},\ \Eprint {https://arxiv.org/abs/2211.14356}
  {arXiv:2211.14356 [hep-ph]} \BibitemShut {NoStop}%
\bibitem [{\citenamefont {Giacalone}\ \emph {et~al.}(2020)\citenamefont
  {Giacalone}, \citenamefont {Schenke},\ and\ \citenamefont
  {Shen}}]{Giacalone:2020byk}%
  \BibitemOpen
  \bibfield  {author} {\bibinfo {author} {\bibfnamefont {G.}~\bibnamefont
  {Giacalone}}, \bibinfo {author} {\bibfnamefont {B.}~\bibnamefont {Schenke}},\
  and\ \bibinfo {author} {\bibfnamefont {C.}~\bibnamefont {Shen}},\ }\bibfield
  {title} {\bibinfo {title} {{Observable signatures of initial state momentum
  anisotropies in nuclear collisions}},\ }\href
  {https://doi.org/10.1103/PhysRevLett.125.192301} {\bibfield  {journal}
  {\bibinfo  {journal} {Phys. Rev. Lett.}\ }\textbf {\bibinfo {volume} {125}},\
  \bibinfo {pages} {192301} (\bibinfo {year} {2020})},\ \Eprint
  {https://arxiv.org/abs/2006.15721} {arXiv:2006.15721 [nucl-th]} \BibitemShut
  {NoStop}%
\bibitem [{\citenamefont {Sievert}\ and\ \citenamefont
  {Noronha-Hostler}(2019)}]{Sievert:2019zjr}%
  \BibitemOpen
  \bibfield  {author} {\bibinfo {author} {\bibfnamefont {M.~D.}\ \bibnamefont
  {Sievert}}\ and\ \bibinfo {author} {\bibfnamefont {J.}~\bibnamefont
  {Noronha-Hostler}},\ }\bibfield  {title} {\bibinfo {title} {{CERN Large
  Hadron Collider system size scan predictions for PbPb, XeXe, ArAr, and OO
  with relativistic hydrodynamics}},\ }\href
  {https://doi.org/10.1103/PhysRevC.100.024904} {\bibfield  {journal} {\bibinfo
   {journal} {Phys. Rev. C}\ }\textbf {\bibinfo {volume} {100}},\ \bibinfo
  {pages} {024904} (\bibinfo {year} {2019})},\ \Eprint
  {https://arxiv.org/abs/1901.01319} {arXiv:1901.01319 [nucl-th]} \BibitemShut
  {NoStop}%
\bibitem [{\citenamefont {Park}\ \emph {et~al.}(2017)\citenamefont {Park},
  \citenamefont {Shen}, \citenamefont {Jeon},\ and\ \citenamefont
  {Gale}}]{Park:2016jap}%
  \BibitemOpen
  \bibfield  {author} {\bibinfo {author} {\bibfnamefont {C.}~\bibnamefont
  {Park}}, \bibinfo {author} {\bibfnamefont {C.}~\bibnamefont {Shen}}, \bibinfo
  {author} {\bibfnamefont {S.}~\bibnamefont {Jeon}},\ and\ \bibinfo {author}
  {\bibfnamefont {C.}~\bibnamefont {Gale}},\ }\bibfield  {title} {\bibinfo
  {title} {{Rapidity-dependent jet energy loss in small systems with
  finite-size effects and running coupling}},\ }\href
  {https://doi.org/10.1016/j.nuclphysbps.2017.05.066} {\bibfield  {journal}
  {\bibinfo  {journal} {Nucl. Part. Phys. Proc.}\ }\textbf {\bibinfo {volume}
  {289-290}},\ \bibinfo {pages} {289} (\bibinfo {year} {2017})},\ \Eprint
  {https://arxiv.org/abs/1612.06754} {arXiv:1612.06754 [nucl-th]} \BibitemShut
  {NoStop}%
\bibitem [{\citenamefont {Kolbe}\ and\ \citenamefont
  {Horowitz}(2019)}]{Kolbe:2015rvk}%
  \BibitemOpen
  \bibfield  {author} {\bibinfo {author} {\bibfnamefont {I.}~\bibnamefont
  {Kolbe}}\ and\ \bibinfo {author} {\bibfnamefont {W.~A.}\ \bibnamefont
  {Horowitz}},\ }\bibfield  {title} {\bibinfo {title} {{Short path length
  corrections to Djordjevic-Gyulassy-Levai-Vitev energy loss}},\ }\href
  {https://doi.org/10.1103/PhysRevC.100.024913} {\bibfield  {journal} {\bibinfo
   {journal} {Phys. Rev. C}\ }\textbf {\bibinfo {volume} {100}},\ \bibinfo
  {pages} {024913} (\bibinfo {year} {2019})},\ \Eprint
  {https://arxiv.org/abs/1511.09313} {arXiv:1511.09313 [hep-ph]} \BibitemShut
  {NoStop}%
\bibitem [{\citenamefont {Adam}\ \emph {et~al.}(2015)\citenamefont {Adam} \emph
  {et~al.}}]{ALICE:2014xsp}%
  \BibitemOpen
  \bibfield  {author} {\bibinfo {author} {\bibfnamefont {J.}~\bibnamefont
  {Adam}} \emph {et~al.} (\bibinfo {collaboration} {ALICE}),\ }\bibfield
  {title} {\bibinfo {title} {{Centrality dependence of particle production in
  p-Pb collisions at $\sqrt{s_{\rm NN} }$= 5.02 TeV}},\ }\href
  {https://doi.org/10.1103/PhysRevC.91.064905} {\bibfield  {journal} {\bibinfo
  {journal} {Phys. Rev. C}\ }\textbf {\bibinfo {volume} {91}},\ \bibinfo
  {pages} {064905} (\bibinfo {year} {2015})},\ \Eprint
  {https://arxiv.org/abs/1412.6828} {arXiv:1412.6828 [nucl-ex]} \BibitemShut
  {NoStop}%
\bibitem [{\citenamefont {Brewer}\ \emph {et~al.}(2022)\citenamefont {Brewer},
  \citenamefont {Brodsky},\ and\ \citenamefont {Rajagopal}}]{Brewer:2021hmh}%
  \BibitemOpen
  \bibfield  {author} {\bibinfo {author} {\bibfnamefont {J.}~\bibnamefont
  {Brewer}}, \bibinfo {author} {\bibfnamefont {Q.}~\bibnamefont {Brodsky}},\
  and\ \bibinfo {author} {\bibfnamefont {K.}~\bibnamefont {Rajagopal}},\
  }\bibfield  {title} {\bibinfo {title} {{Disentangling jet modification in jet
  simulations and in Z+jet data}},\ }\href
  {https://doi.org/10.1007/JHEP02(2022)175} {\bibfield  {journal} {\bibinfo
  {journal} {JHEP}\ }\textbf {\bibinfo {volume} {02}},\ \bibinfo {pages}
  {175}},\ \Eprint {https://arxiv.org/abs/2110.13159} {arXiv:2110.13159
  [hep-ph]} \BibitemShut {NoStop}%
\bibitem [{\citenamefont {Brewer}\ \emph {et~al.}(2019)\citenamefont {Brewer},
  \citenamefont {Milhano},\ and\ \citenamefont {Thaler}}]{Brewer:2018dfs}%
  \BibitemOpen
  \bibfield  {author} {\bibinfo {author} {\bibfnamefont {J.}~\bibnamefont
  {Brewer}}, \bibinfo {author} {\bibfnamefont {J.~G.}\ \bibnamefont
  {Milhano}},\ and\ \bibinfo {author} {\bibfnamefont {J.}~\bibnamefont
  {Thaler}},\ }\bibfield  {title} {\bibinfo {title} {{Sorting out quenched
  jets}},\ }\href {https://doi.org/10.1103/PhysRevLett.122.222301} {\bibfield
  {journal} {\bibinfo  {journal} {Phys. Rev. Lett.}\ }\textbf {\bibinfo
  {volume} {122}},\ \bibinfo {pages} {222301} (\bibinfo {year} {2019})},\
  \Eprint {https://arxiv.org/abs/1812.05111} {arXiv:1812.05111 [hep-ph]}
  \BibitemShut {NoStop}%
\bibitem [{\citenamefont {Marzani}\ \emph {et~al.}(2019)\citenamefont
  {Marzani}, \citenamefont {Soyez},\ and\ \citenamefont
  {Spannowsky}}]{Marzani:2019hun}%
  \BibitemOpen
  \bibfield  {author} {\bibinfo {author} {\bibfnamefont {S.}~\bibnamefont
  {Marzani}}, \bibinfo {author} {\bibfnamefont {G.}~\bibnamefont {Soyez}},\
  and\ \bibinfo {author} {\bibfnamefont {M.}~\bibnamefont {Spannowsky}},\
  }\href {https://doi.org/10.1007/978-3-030-15709-8} {\emph {\bibinfo {title}
  {{Looking inside jets: an introduction to jet substructure and boosted-object
  phenomenology}}}},\ Vol.\ \bibinfo {volume} {958}\ (\bibinfo  {publisher}
  {Springer},\ \bibinfo {year} {2019})\ \Eprint
  {https://arxiv.org/abs/1901.10342} {arXiv:1901.10342 [hep-ph]} \BibitemShut
  {NoStop}%
\bibitem [{\citenamefont {Larkoski}\ \emph {et~al.}(2020)\citenamefont
  {Larkoski}, \citenamefont {Moult},\ and\ \citenamefont
  {Nachman}}]{Larkoski:2017jix}%
  \BibitemOpen
  \bibfield  {author} {\bibinfo {author} {\bibfnamefont {A.~J.}\ \bibnamefont
  {Larkoski}}, \bibinfo {author} {\bibfnamefont {I.}~\bibnamefont {Moult}},\
  and\ \bibinfo {author} {\bibfnamefont {B.}~\bibnamefont {Nachman}},\
  }\bibfield  {title} {\bibinfo {title} {{Jet Substructure at the Large Hadron
  Collider: A Review of Recent Advances in Theory and Machine Learning}},\
  }\href {https://doi.org/10.1016/j.physrep.2019.11.001} {\bibfield  {journal}
  {\bibinfo  {journal} {Phys. Rept.}\ }\textbf {\bibinfo {volume} {841}},\
  \bibinfo {pages} {1} (\bibinfo {year} {2020})},\ \Eprint
  {https://arxiv.org/abs/1709.04464} {arXiv:1709.04464 [hep-ph]} \BibitemShut
  {NoStop}%
\bibitem [{\citenamefont {Kogler}\ \emph {et~al.}(2019)\citenamefont {Kogler}
  \emph {et~al.}}]{Kogler:2018hem}%
  \BibitemOpen
  \bibfield  {author} {\bibinfo {author} {\bibfnamefont {R.}~\bibnamefont
  {Kogler}} \emph {et~al.},\ }\bibfield  {title} {\bibinfo {title} {{Jet
  Substructure at the Large Hadron Collider: Experimental Review}},\ }\href
  {https://doi.org/10.1103/RevModPhys.91.045003} {\bibfield  {journal}
  {\bibinfo  {journal} {Rev. Mod. Phys.}\ }\textbf {\bibinfo {volume} {91}},\
  \bibinfo {pages} {045003} (\bibinfo {year} {2019})},\ \Eprint
  {https://arxiv.org/abs/1803.06991} {arXiv:1803.06991 [hep-ex]} \BibitemShut
  {NoStop}%
\bibitem [{\citenamefont {Casalderrey-Solana}\ \emph
  {et~al.}(2020)\citenamefont {Casalderrey-Solana}, \citenamefont {Milhano},
  \citenamefont {Pablos},\ and\ \citenamefont
  {Rajagopal}}]{Casalderrey-Solana:2019ubu}%
  \BibitemOpen
  \bibfield  {author} {\bibinfo {author} {\bibfnamefont {J.}~\bibnamefont
  {Casalderrey-Solana}}, \bibinfo {author} {\bibfnamefont {G.}~\bibnamefont
  {Milhano}}, \bibinfo {author} {\bibfnamefont {D.}~\bibnamefont {Pablos}},\
  and\ \bibinfo {author} {\bibfnamefont {K.}~\bibnamefont {Rajagopal}},\
  }\bibfield  {title} {\bibinfo {title} {{Modification of Jet Substructure in
  Heavy Ion Collisions as a Probe of the Resolution Length of Quark-Gluon
  Plasma}},\ }\href {https://doi.org/10.1007/JHEP01(2020)044} {\bibfield
  {journal} {\bibinfo  {journal} {JHEP}\ }\textbf {\bibinfo {volume} {01}},\
  \bibinfo {pages} {044}},\ \Eprint {https://arxiv.org/abs/1907.11248}
  {arXiv:1907.11248 [hep-ph]} \BibitemShut {NoStop}%
\bibitem [{\citenamefont {Mulligan}\ and\ \citenamefont
  {Ploskon}(2020)}]{Mulligan:2020tim}%
  \BibitemOpen
  \bibfield  {author} {\bibinfo {author} {\bibfnamefont {J.}~\bibnamefont
  {Mulligan}}\ and\ \bibinfo {author} {\bibfnamefont {M.}~\bibnamefont
  {Ploskon}},\ }\bibfield  {title} {\bibinfo {title} {{Identifying groomed jet
  splittings in heavy-ion collisions}},\ }\href
  {https://doi.org/10.1103/PhysRevC.102.044913} {\bibfield  {journal} {\bibinfo
   {journal} {Phys. Rev. C}\ }\textbf {\bibinfo {volume} {102}},\ \bibinfo
  {pages} {044913} (\bibinfo {year} {2020})},\ \Eprint
  {https://arxiv.org/abs/2006.01812} {arXiv:2006.01812 [hep-ph]} \BibitemShut
  {NoStop}%
\bibitem [{\citenamefont {Apolin\'ario}\ \emph {et~al.}(2018)\citenamefont
  {Apolin\'ario}, \citenamefont {Milhano}, \citenamefont {Ploskon},\ and\
  \citenamefont {Zhang}}]{Apolinario:2017qay}%
  \BibitemOpen
  \bibfield  {author} {\bibinfo {author} {\bibfnamefont {L.}~\bibnamefont
  {Apolin\'ario}}, \bibinfo {author} {\bibfnamefont {J.~G.}\ \bibnamefont
  {Milhano}}, \bibinfo {author} {\bibfnamefont {M.}~\bibnamefont {Ploskon}},\
  and\ \bibinfo {author} {\bibfnamefont {X.}~\bibnamefont {Zhang}},\ }\bibfield
   {title} {\bibinfo {title} {{Novel subjet observables for jet quenching in
  heavy-ion collisions}},\ }\href
  {https://doi.org/10.1140/epjc/s10052-018-5999-2} {\bibfield  {journal}
  {\bibinfo  {journal} {Eur. Phys. J. C}\ }\textbf {\bibinfo {volume} {78}},\
  \bibinfo {pages} {529} (\bibinfo {year} {2018})},\ \Eprint
  {https://arxiv.org/abs/1710.07607} {arXiv:1710.07607 [hep-ph]} \BibitemShut
  {NoStop}%
\bibitem [{\citenamefont {Acharya}\ \emph {et~al.}(2022)\citenamefont {Acharya}
  \emph {et~al.}}]{ALargeIonColliderExperiment:2021mqf}%
  \BibitemOpen
  \bibfield  {author} {\bibinfo {author} {\bibfnamefont {S.}~\bibnamefont
  {Acharya}} \emph {et~al.} (\bibinfo {collaboration} {A Large Ion Collider
  Experiment, ALICE}),\ }\bibfield  {title} {\bibinfo {title} {{Measurement of
  the groomed jet radius and momentum splitting fraction in pp and Pb$-$Pb
  collisions at $\sqrt{s_{NN}} = 5.02$ TeV}},\ }\href
  {https://doi.org/10.1103/PhysRevLett.128.102001} {\bibfield  {journal}
  {\bibinfo  {journal} {Phys. Rev. Lett.}\ }\textbf {\bibinfo {volume} {128}},\
  \bibinfo {pages} {102001} (\bibinfo {year} {2022})},\ \Eprint
  {https://arxiv.org/abs/2107.12984} {arXiv:2107.12984 [nucl-ex]} \BibitemShut
  {NoStop}%
\bibitem [{\citenamefont {Sirunyan}\ \emph {et~al.}(2018)\citenamefont
  {Sirunyan} \emph {et~al.}}]{CMS:2017qlm}%
  \BibitemOpen
  \bibfield  {author} {\bibinfo {author} {\bibfnamefont {A.~M.}\ \bibnamefont
  {Sirunyan}} \emph {et~al.} (\bibinfo {collaboration} {CMS}),\ }\bibfield
  {title} {\bibinfo {title} {{Measurement of the Splitting Function in $pp$ and
  Pb-Pb Collisions at $\sqrt{s_{_{\mathrm{NN}}}} =$ 5.02 TeV}},\ }\href
  {https://doi.org/10.1103/PhysRevLett.120.142302} {\bibfield  {journal}
  {\bibinfo  {journal} {Phys. Rev. Lett.}\ }\textbf {\bibinfo {volume} {120}},\
  \bibinfo {pages} {142302} (\bibinfo {year} {2018})},\ \Eprint
  {https://arxiv.org/abs/1708.09429} {arXiv:1708.09429 [nucl-ex]} \BibitemShut
  {NoStop}%
\bibitem [{\citenamefont {Abdallah}\ \emph {et~al.}(2022)\citenamefont
  {Abdallah} \emph {et~al.}}]{STAR:2021kjt}%
  \BibitemOpen
  \bibfield  {author} {\bibinfo {author} {\bibfnamefont {M.~S.}\ \bibnamefont
  {Abdallah}} \emph {et~al.} (\bibinfo {collaboration} {STAR}),\ }\bibfield
  {title} {\bibinfo {title} {{Differential measurements of jet substructure and
  partonic energy loss in Au+Au collisions at $\sqrt {S_{NN}}$ =200 GeV}},\
  }\href {https://doi.org/10.1103/PhysRevC.105.044906} {\bibfield  {journal}
  {\bibinfo  {journal} {Phys. Rev. C}\ }\textbf {\bibinfo {volume} {105}},\
  \bibinfo {pages} {044906} (\bibinfo {year} {2022})},\ \Eprint
  {https://arxiv.org/abs/2109.09793} {arXiv:2109.09793 [nucl-ex]} \BibitemShut
  {NoStop}%
\bibitem [{\citenamefont {Larkoski}\ \emph {et~al.}(2014)\citenamefont
  {Larkoski}, \citenamefont {Marzani}, \citenamefont {Soyez},\ and\
  \citenamefont {Thaler}}]{Larkoski:2014wba}%
  \BibitemOpen
  \bibfield  {author} {\bibinfo {author} {\bibfnamefont {A.~J.}\ \bibnamefont
  {Larkoski}}, \bibinfo {author} {\bibfnamefont {S.}~\bibnamefont {Marzani}},
  \bibinfo {author} {\bibfnamefont {G.}~\bibnamefont {Soyez}},\ and\ \bibinfo
  {author} {\bibfnamefont {J.}~\bibnamefont {Thaler}},\ }\bibfield  {title}
  {\bibinfo {title} {{Soft Drop}},\ }\href
  {https://doi.org/10.1007/JHEP05(2014)146} {\bibfield  {journal} {\bibinfo
  {journal} {JHEP}\ }\textbf {\bibinfo {volume} {05}},\ \bibinfo {pages}
  {146}},\ \Eprint {https://arxiv.org/abs/1402.2657} {arXiv:1402.2657 [hep-ph]}
  \BibitemShut {NoStop}%
\bibitem [{\citenamefont {Mehtar-Tani}\ \emph {et~al.}(2020)\citenamefont
  {Mehtar-Tani}, \citenamefont {Soto-Ontoso},\ and\ \citenamefont
  {Tywoniuk}}]{Mehtar-Tani:2019rrk}%
  \BibitemOpen
  \bibfield  {author} {\bibinfo {author} {\bibfnamefont {Y.}~\bibnamefont
  {Mehtar-Tani}}, \bibinfo {author} {\bibfnamefont {A.}~\bibnamefont
  {Soto-Ontoso}},\ and\ \bibinfo {author} {\bibfnamefont {K.}~\bibnamefont
  {Tywoniuk}},\ }\bibfield  {title} {\bibinfo {title} {{Dynamical grooming of
  QCD jets}},\ }\href {https://doi.org/10.1103/PhysRevD.101.034004} {\bibfield
  {journal} {\bibinfo  {journal} {Phys. Rev. D}\ }\textbf {\bibinfo {volume}
  {101}},\ \bibinfo {pages} {034004} (\bibinfo {year} {2020})},\ \Eprint
  {https://arxiv.org/abs/1911.00375} {arXiv:1911.00375 [hep-ph]} \BibitemShut
  {NoStop}%
\bibitem [{\citenamefont {Dokshitzer}\ \emph {et~al.}(1997)\citenamefont
  {Dokshitzer}, \citenamefont {Leder}, \citenamefont {Moretti},\ and\
  \citenamefont {Webber}}]{Dokshitzer:1997in}%
  \BibitemOpen
  \bibfield  {author} {\bibinfo {author} {\bibfnamefont {Y.~L.}\ \bibnamefont
  {Dokshitzer}}, \bibinfo {author} {\bibfnamefont {G.~D.}\ \bibnamefont
  {Leder}}, \bibinfo {author} {\bibfnamefont {S.}~\bibnamefont {Moretti}},\
  and\ \bibinfo {author} {\bibfnamefont {B.~R.}\ \bibnamefont {Webber}},\
  }\bibfield  {title} {\bibinfo {title} {{Better jet clustering algorithms}},\
  }\href {https://doi.org/10.1088/1126-6708/1997/08/001} {\bibfield  {journal}
  {\bibinfo  {journal} {JHEP}\ }\textbf {\bibinfo {volume} {08}},\ \bibinfo
  {pages} {001}},\ \Eprint {https://arxiv.org/abs/hep-ph/9707323}
  {arXiv:hep-ph/9707323} \BibitemShut {NoStop}%
\bibitem [{\citenamefont {Schenke}\ \emph {et~al.}(2020)\citenamefont
  {Schenke}, \citenamefont {Shen},\ and\ \citenamefont
  {Tribedy}}]{Schenke:2020mbo}%
  \BibitemOpen
  \bibfield  {author} {\bibinfo {author} {\bibfnamefont {B.}~\bibnamefont
  {Schenke}}, \bibinfo {author} {\bibfnamefont {C.}~\bibnamefont {Shen}},\ and\
  \bibinfo {author} {\bibfnamefont {P.}~\bibnamefont {Tribedy}},\ }\bibfield
  {title} {\bibinfo {title} {{Running the gamut of high energy nuclear
  collisions}},\ }\href {https://doi.org/10.1103/PhysRevC.102.044905}
  {\bibfield  {journal} {\bibinfo  {journal} {Phys. Rev. C}\ }\textbf {\bibinfo
  {volume} {102}},\ \bibinfo {pages} {044905} (\bibinfo {year} {2020})},\
  \Eprint {https://arxiv.org/abs/2005.14682} {arXiv:2005.14682 [nucl-th]}
  \BibitemShut {NoStop}%
\bibitem [{\citenamefont {Adare}\ \emph {et~al.}(2015)\citenamefont {Adare}
  \emph {et~al.}}]{PHENIX:2015idk}%
  \BibitemOpen
  \bibfield  {author} {\bibinfo {author} {\bibfnamefont {A.}~\bibnamefont
  {Adare}} \emph {et~al.} (\bibinfo {collaboration} {PHENIX}),\ }\bibfield
  {title} {\bibinfo {title} {{Measurements of elliptic and triangular flow in
  high-multiplicity $^{3}$He$+$Au collisions at $\sqrt{s_{_{NN}}}=200$ GeV}},\
  }\href {https://doi.org/10.1103/PhysRevLett.115.142301} {\bibfield  {journal}
  {\bibinfo  {journal} {Phys. Rev. Lett.}\ }\textbf {\bibinfo {volume} {115}},\
  \bibinfo {pages} {142301} (\bibinfo {year} {2015})},\ \Eprint
  {https://arxiv.org/abs/1507.06273} {arXiv:1507.06273 [nucl-ex]} \BibitemShut
  {NoStop}%
\bibitem [{\citenamefont {Dulat}\ \emph {et~al.}(2016)\citenamefont {Dulat},
  \citenamefont {Hou}, \citenamefont {Gao}, \citenamefont {Guzzi},
  \citenamefont {Huston}, \citenamefont {Nadolsky}, \citenamefont {Pumplin},
  \citenamefont {Schmidt}, \citenamefont {Stump},\ and\ \citenamefont
  {Yuan}}]{Dulat:2015mca}%
  \BibitemOpen
  \bibfield  {author} {\bibinfo {author} {\bibfnamefont {S.}~\bibnamefont
  {Dulat}}, \bibinfo {author} {\bibfnamefont {T.-J.}\ \bibnamefont {Hou}},
  \bibinfo {author} {\bibfnamefont {J.}~\bibnamefont {Gao}}, \bibinfo {author}
  {\bibfnamefont {M.}~\bibnamefont {Guzzi}}, \bibinfo {author} {\bibfnamefont
  {J.}~\bibnamefont {Huston}}, \bibinfo {author} {\bibfnamefont
  {P.}~\bibnamefont {Nadolsky}}, \bibinfo {author} {\bibfnamefont
  {J.}~\bibnamefont {Pumplin}}, \bibinfo {author} {\bibfnamefont
  {C.}~\bibnamefont {Schmidt}}, \bibinfo {author} {\bibfnamefont
  {D.}~\bibnamefont {Stump}},\ and\ \bibinfo {author} {\bibfnamefont {C.~P.}\
  \bibnamefont {Yuan}},\ }\bibfield  {title} {\bibinfo {title} {{New parton
  distribution functions from a global analysis of quantum chromodynamics}},\
  }\href {https://doi.org/10.1103/PhysRevD.93.033006} {\bibfield  {journal}
  {\bibinfo  {journal} {Phys. Rev. D}\ }\textbf {\bibinfo {volume} {93}},\
  \bibinfo {pages} {033006} (\bibinfo {year} {2016})},\ \Eprint
  {https://arxiv.org/abs/1506.07443} {arXiv:1506.07443 [hep-ph]} \BibitemShut
  {NoStop}%
\bibitem [{\citenamefont {Eskola}\ \emph {et~al.}(2022)\citenamefont {Eskola},
  \citenamefont {Paakkinen}, \citenamefont {Paukkunen},\ and\ \citenamefont
  {Salgado}}]{Eskola:2021nhw}%
  \BibitemOpen
  \bibfield  {author} {\bibinfo {author} {\bibfnamefont {K.~J.}\ \bibnamefont
  {Eskola}}, \bibinfo {author} {\bibfnamefont {P.}~\bibnamefont {Paakkinen}},
  \bibinfo {author} {\bibfnamefont {H.}~\bibnamefont {Paukkunen}},\ and\
  \bibinfo {author} {\bibfnamefont {C.~A.}\ \bibnamefont {Salgado}},\
  }\bibfield  {title} {\bibinfo {title} {{EPPS21: a global QCD analysis of
  nuclear PDFs}},\ }\href {https://doi.org/10.1140/epjc/s10052-022-10359-0}
  {\bibfield  {journal} {\bibinfo  {journal} {Eur. Phys. J. C}\ }\textbf
  {\bibinfo {volume} {82}},\ \bibinfo {pages} {413} (\bibinfo {year} {2022})},\
  \Eprint {https://arxiv.org/abs/2112.12462} {arXiv:2112.12462 [hep-ph]}
  \BibitemShut {NoStop}%
\bibitem [{\citenamefont {Hou}\ \emph {et~al.}(2021)\citenamefont {Hou} \emph
  {et~al.}}]{Hou:2019efy}%
  \BibitemOpen
  \bibfield  {author} {\bibinfo {author} {\bibfnamefont {T.-J.}\ \bibnamefont
  {Hou}} \emph {et~al.},\ }\bibfield  {title} {\bibinfo {title} {{New CTEQ
  global analysis of quantum chromodynamics with high-precision data from the
  LHC}},\ }\href {https://doi.org/10.1103/PhysRevD.103.014013} {\bibfield
  {journal} {\bibinfo  {journal} {Phys. Rev. D}\ }\textbf {\bibinfo {volume}
  {103}},\ \bibinfo {pages} {014013} (\bibinfo {year} {2021})},\ \Eprint
  {https://arxiv.org/abs/1912.10053} {arXiv:1912.10053 [hep-ph]} \BibitemShut
  {NoStop}%
\end{thebibliography}%
    
\end{document}